\global\def\draftcontrol{0}

   \def\versionno{ n4rn3}

\catcode`\@=11

\expandafter\ifx\csname draftcontrol\endcsname\relax\global\def\draftcontrol{0}
\fi

{\count255=\time\divide\count255 by 60
\xdef\hourmin{\number\count255}
\multiply\count255 by-60\advance\count255 by\time
\xdef\hourmin{\hourmin:\ifnum\count255<10 0\fi\the\count255}}
\def\draftdate{\number\month/\number\day/\number\year\ \ \ \hourmin }

\newcommand\makepapertitle{\par
  \begingroup
    \renewcommand\thefootnote{\@fnsymbol\c@footnote}%
    \def\@makefnmark{\rlap{\@textsuperscript{\normalfont\@thefnmark}}}%
    \long\def\@makefntext##1{\parindent 1em\noindent
            \hb@xt@1.8em{%
                \hss\@textsuperscript{\normalfont\@thefnmark}}##1}%
     \newpage
     \global\@topnum\z@   
     \@makepapertitle
     \thispagestyle{empty}\@thanks
  \endgroup
  \setcounter{footnote}{0}%
  \global\let\thanks\relax
  \global\let\makepapertitle\relax
  \global\let\@makepapertitle\relax
  \global\let\@thanks\@empty
  \global\let\@author\@empty
  \global\let\@date\@empty
  \global\let\@title\@empty
  \global\let\title\relax
  \global\let\author\relax
  \global\let\date\relax
  \global\let\and\relax
  \def\version{\let\version\@version\@gobble}
}
\def\@makepapertitle{%
  \newpage
   \ifnum\draftcontrol=1 {}
   \version\versionno
   \vskip 3em%
   \else
   \hfill\hbox to 3cm {\parbox{4cm}{\@pubnum}\hss}%
   \vskip 3em%
   \fi
   \begin{center}%
   \let \footnote \thanks
     {\LARGE {\@title}}%
     \vskip 1.5em%
     {\normalsize
       \lineskip .5em%
       \begin{tabular}[t]{c}%
         \@author
       \end{tabular}\par}%
     \vskip 1.5em%
     {\@bstract}%
     \end{center}%
     \vskip 1.5em
     \@date%
   \par
}

\gdef\@pubnum{}
\def\pubnum#1{%
  \gdef\@pubnum{#1}}

\gdef\@bstract{}
\def\Abstract#1{%
  \gdef\@bstract{%
   \parbox{\textwidth-0pc}{%
   \centerline{\bf Abstract}\penalty1000%
\kern.2cm%
\noindent
\renewcommand\baselinestretch{1.0}%
{#1}}}
}

\def\ps@paper{\let\@mkboth\@gobbletwo%
     \ifnum\draftcontrol=1
    \def\@oddfoot{\hbox to \textwidth{\tiny \versionno \hfil\tiny\draftdate}%
    \hskip -\textwidth \hbox to \textwidth{\hfil\rm\thepage\hfil}}%
     \else\def\@oddfoot{\hbox to \textwidth{\hfil\rm\thepage\hfil}}
     \fi
     \let\@evenfoot\@oddfoot
}

\def\body{\clearpage
          \pagestyle{paper}
    }

\def\@version#1{\ifnum\draftcontrol=1
\typeout{}\typeout{#1}\typeout{}
\vskip3mm\centerline{\hbox{\fbox{\normalsize{\tt DRAFT -- #1 -- }
                   {\draftdate}}}}\vskip3mm
\fi}
\let\version\@version
\long\def\eqlabel#1{\ifnum\draftcontrol=1
                    \tag@false  
                    \tag*{(\theequation) \hbox to -0.2cm{\hspace{0cm}\small{#1}\hss}}
                    \refstepcounter{equation}
                    \edef\@currentlabel{\theequation}
                    \ltx@label{#1}          
                    \else
                    \label{#1}
                    \fi
                    }
\let\st@bibitem\@bibitem
\let\st@lbibitem\@lbibitem
\ifnum\draftcontrol=1
  \def\@bibitem#1{%
    \st@bibitem{#1}\a@@label{#1}\ignorespaces}
  \def\@lbibitem[#1]#2{%
    \st@lbibitem[#1]{#2}\a@@label{#2}\ignorespaces}
  \def\a@@label#1{%
    \gdef\a@lab{\smash{\normalfont\small#1}}
    \ifvmode
      \if@inlabel
        \global\setbox\@labels\hbox{%
          \llap{\a@lab\let\a@lab\relax
                \kern\@totalleftmargin\kern\marginparsep}%
          \box\@labels}%
      \fi
    \fi}
\fi

\documentclass[12pt,letterpaper]{article}

\usepackage{amsmath,amssymb,array,calc,epsfig,rotating,bm}
\usepackage[sort]{cite}
\usepackage{graphicx}
\usepackage{psfrag,verbatim}
\usepackage{xcolor}
\usepackage{hyperref}

\ifnum\draftcontrol=1
\fi

\renewcommand\baselinestretch{1.25}
\renewcommand\section{\@startsection {section}{1}{\z@}%
                                   {-3.5ex \@plus -1ex \@minus -.2ex}%
                                   {2.3ex \@plus.2ex}%
                                   {\normalfont\large\bfseries}}
\renewcommand\subsection{\@startsection{subsection}{2}{\z@}%
                                   {-3.25ex\@plus -1ex \@minus -.2ex}%
                                   {1.5ex \@plus .2ex}%
                                   {\normalfont\normalsize\bfseries}}
\renewcommand\subsubsection{\@startsection{subsubsection}{3}{\z@}%
                                   {-3.25ex\@plus -1ex \@minus -.2ex}%
                                   {1.5ex \@plus .2ex}%
                                   {\normalfont\normalsize\it}}
\renewcommand\paragraph{\@startsection{paragraph}{4}{\z@}%
                                   {-3.25ex\@plus -1ex \@minus -.2ex}%
                                   {1.5ex \@plus .2ex}%
                                   {\normalfont\normalsize\bf}}

\numberwithin{equation}{section}

\def\revise#1       {\raisebox{-0em}{\rule{3pt}{1em}}%
                     \marginpar{\raisebox{.5em}{\vrule width3pt\
                     \vrule width0pt height 0pt depth0.5em
                     \hbox to 0cm{\hspace{0cm}{%
                     \parbox[t]{4em}{\raggedright\footnotesize{#1}}}\hss}}}}

\newcommand{\ie}{{\it i.e.,}\ }

\def\calm         {{\cal M}}
\def\caln         {{\cal N}}

\def\calp         {{\cal P}}

\def\del          {\partial}

\def\sqr#1#2{{\vcenter{\vbox{\hrule height.#2pt
 \hbox{\vrule width.#2pt height#1pt \kern#1pt
 \vrule width.#2pt}\hrule height.#2pt}}}}

\def\aa1{\phi}
\def\cc1{\psi}

\catcode`\@=12

\begin{document}


\title{\bf On the relevance of GIKS instability of charged $\caln=4$ SYM plasma}

\date{February 16, 2025}

\author{
Alex Buchel\\[0.4cm]
\it Department of Physics and Astronomy\\ 
\it University of Western Ontario\\
\it London, Ontario N6A 5B7, Canada\\
\it Perimeter Institute for Theoretical Physics\\
\it Waterloo, Ontario N2J 2W9, Canada\\
}

\Abstract{Recently Gladden, Ivo, Kovtun, and Starinets (GIKS) identified a novel
instability of ${\cal N}=4$ supersymmetric Yang-Mills plasma with a
diagonal $U(1)$ $R$-symmetry chemical potential.  Since ${\cal N}=4$
SYM plasma has $R$-charged matter fields, it might suffer from the
standard superconducting instability. We show that while the
superconducting instability is indeed present, it happens at a lower
critical temperature compare to that of the GIKS instability.}

\makepapertitle

\body

\version\versionno


Strongly coupled $\caln=4$ $SU(N_c)$ supersymmetric Yang-Mills (SYM) theory
has a holographic dual as type IIB string theory in $AdS_5\times S^5$ \cite{Maldacena:1997re}.
In the planar limit, and at large $N_c$, the thermal states of the SYM are represented
by black branes in type IIB supergravity. This gauge theory has $SU(4)$ $R$-symmetry,
and one can study its charged plasma, with the same chemical chemical potential $\mu$
for all $U(1)$ factors of the maximal Abelian subgroup $U(1)^3\subset SU(4)$.
In this case the gravitational dual is represented by a
Reissner-Nordstrom (RN) black brane in asymptotically $AdS_5$ space-time.
One reason why this model is interesting, is that it allows to
study extremal horizons (where it is possible to take a limit of vanishing temperature)
in the best explored holographic example.

In \cite{Gladden:2024ssb} (GIKS) the authors argued that
the extremal limit is never achieved\footnote{Identical phenomenon
occurs in other holographic models as well \cite{Buchel:2025jup}.}: 
there is a critical temperature, more precisely the critical value of $\frac{T}{\mu}$,
below which the charged SYM plasma is unstable to fluctuations of
the off-diagonal gauge fields and the neutral scalars
of the bulk gravitational description, when interpreted  within
STU consistent truncation \cite{Behrndt:1998jd,Cvetic:1999xp,Cvetic:2000nc} of type IIB supergravity. 
However, $\caln=4$ SYM also has $R$-charged matter fields, which could condense at sufficiently
low values of $\frac T\mu$ \cite{Gubser:2009qm}. STU truncation does not provide
gravitational description of such charged matter. 

The question we address in this note is whether the GIKS instability
is physically relevant: \ie whether the standard superconducting instability in charged SYM plasma
occurs before the former can be triggered. Jumping to the result, we find that 
\begin{equation}
\frac{T_c}{\mu}\bigg|_{\rm superconducting}\ =\ 0.57186(4)\ \frac{T_c}{\mu}\bigg|_{\rm GIKS}\,,
\eqlabel{tcrit}
\end{equation}
\ie the superconducting instability actually occurs at a lower temperature compare
to the onset of the instability reported in \cite{Gladden:2024ssb}.
Our result highlights the importance of understanding the
end-point of the linearized GIKS instability.

The rest of this note explains how we reached the conclusion \eqref{tcrit}.
As we already alluded to, it is important to enlarge the dual gravitational description
beyond STU. Luckily, such a model is known \cite{Bobev:2010de}:
\begin{equation}
\begin{split}
&S_5=\frac{1}{4\pi G_5}\int_{\calm_5}d^5\xi\sqrt{-g} \biggl[
\frac R4-\frac 14 \biggl(\rho^4 \nu^{-4} F_{\mu\nu}^{(1)}F^{(1)\mu\nu}+\rho^4\nu^4F_{\mu\nu}^{(2)}
F^{(2)\mu\nu}+\rho^{-8}F_{\mu\nu}^{(3)}F^{(3)\mu\nu}\biggr)\\
&-\frac 12 \sum_{j=1}^4 \left(\del_\mu\phi_j\right)^2-3\left(\del_\mu\alpha\right)^2
-\left(\del_\mu\beta\right)^2-\frac 18 \sinh^2(2\phi_1)\left(
\del_\mu\theta_1+\left(A_\mu^{(1)}+A_\mu^{(2)}-A_\mu^{(3)}\right)
\right)^2\\
&-\frac 18 \sinh^2(2\phi_2)\left(
\del_\mu\theta_2+\left(A_\mu^{(1)}-A_\mu^{(2)}+A_\mu^{(3)}\right)\right)^2
-\frac 18 \sinh^2(2\phi_3)(
\del_\mu\theta_3+(-A_\mu^{(1)}+A_\mu^{(2)}\\
&+A_\mu^{(3)}))^2-\frac 18 \sinh^2(2\phi_4)\left(
\del_\mu\theta_4-\left(A_\mu^{(1)}+A_\mu^{(2)}+A_\mu^{(3)}\right)\right)^2
-\calp
\biggr]\,.
\end{split}
\eqlabel{model1}
\end{equation}
Here the $F^{(J)}$ are the field strengths of the $U(1)$ gauge fields $A^{(J)}$
(dual
to $U(1)^3\subset SU(4)$ global $R$-symmetry currents),
$\alpha$ and $\beta$ are the STU-truncation bulk scalars, $\phi_i$
are the bulk scalars dual to dimension-3 fermion bilinears of the boundary SYM,
and $\calp$ is the scalar potential. We introduced
\begin{equation}
\rho\equiv e^\alpha\,,\qquad \nu\equiv e^\beta\,.
\eqlabel{model2}
\end{equation}
The scalar potential, $\calp$, is given in terms of a superpotential
\begin{equation}
\calp=\frac{g^2}{8}\biggl[\
\sum_{j=1}^4 \left(\frac{\del W}{\del\phi_j}\right)^2+\frac 16
\left(\frac{\del W}{\del\alpha}\right)^2
+\frac 12\left(\frac{\del W}{\del \beta}\right)^2\
\biggr]-\frac{g^2}{3} W^2\,,
\eqlabel{model3}
\end{equation}
where
\begin{equation}
\begin{split}
W=&-\frac{1}{4\rho^2\nu^2}\biggl[
\left(1+\nu^4-\nu^2\rho^6\right)\cosh(2\phi_1)+\left(-1+\nu^4+\nu^2\rho^6\right)
\cosh(2\phi_2)\\
&+\left(1-\nu^4+\nu^2\rho^6\right)\cosh(2\phi_3)+
\left(1+\nu^4+\nu^2\rho^6\right)\cosh(2\phi_4)
\biggr]\,.
\end{split}
\eqlabel{model4}
\end{equation}
In what follows we set the gauged supergravity coupling $g=1$, this corresponds to setting the
asymptotic $AdS_5$ radius to $L=2$. The five dimensional gravitational constant
$G_5$ is related to the rank of the supersymmetric $\caln=4$ $SU(N_c)$ UV fixed point as
\begin{equation}
G_5=\frac{4\pi}{N^2}\,.
\eqlabel{defg5}
\end{equation}
Different consistent truncations of \eqref{model4} were discussed in \cite{Buchel:2020vkv}.
For this paper we care
about\footnote{While we do not present the details here, we did verify
that other then the instabilities of {\bf (Tr-II)} and {\bf (Tr-III)}
discussed below, the RN black branes of {\bf (Tr-I)} do not have
additional spatially homogeneous and isotropic instabilities, when described within
the gravitational dual \eqref{model1}.}:
\begin{itemize}
\item {\bf(Tr-I)},  the minimal holographic dual to the SYM with a diagonal $U(1)$ $R$-symmetry is
\begin{equation}
\alpha=\beta=\phi_j=\theta_j=0\,,\qquad A_\mu^{(1)}=A_\mu^{(2)}=A_\mu^{(3)}\,;
\eqlabel{min}
\end{equation}
\item {\bf(Tr-II)}, the STU model,
\begin{equation}
\phi_j=\theta_j=0\,;
\eqlabel{stu}
\end{equation}
\item {\bf(Tr-III)}, the truncation  describing the superconducting phase of the charged
$\caln=4$ SYM plasma,
\begin{equation}
\alpha=\beta=\phi_{1,2,3}=\theta_{1,2,3}=0\,,\qquad A_\mu^{(1)}=A_\mu^{(2)}=A_\mu^{(3)}\,.
\eqlabel{super}
\end{equation}
\end{itemize}

\underline{{\bf (Tr-I):}} The RN black brane is a solution of the effective action \eqref{model1}
with  \eqref{min}:
\begin{equation}
\begin{split}
&ds_5=\frac 4r\biggl(- f\ dt^2+d{\bm x}^2\biggr)+\frac{dr^2}{r^2 f}\,,\qquad f=1-(p^2+1)\ r^2+p^2\ r^3 \\
&A_\mu^{(j)}=p\ (1-r)\ \delta^t_{\mu}\,,
\end{split}
\eqlabel{rn}
\end{equation}
where the radial coordinate $r\in (0,1]$ (from the asymptotic $AdS_5$ boundary to the
regular Schwarzschild horizon), and parameter $p$ determines
\begin{equation}
\frac{T}{\mu}=\frac{2-p^2}{2\pi p}\,,
\eqlabel{tc2}
\end{equation}
with the extremal limit achieved as $p\to \sqrt{2}$.

\underline{{\bf (Tr-II):}} The RN black brane \eqref{rn}, is unstable in grand canonical
ensemble, within the STU model \eqref{stu}, to an ordered phase 
\cite{Buchel:2025cve,Buchel:2025tjq} below\footnote{See \cite{Henriksson:2019zph,Anabalon:2024lgp}
for earlier related work.}
\begin{equation}
\frac{T_c}{\mu}=\frac{1}{2\pi}\ \equiv \frac{T_c}{\mu}\bigg|_{\rm GIKS}\,,
\eqlabel{tc1}
\end{equation}
as it turns out that \eqref{tc1} is also the onset of the GIKS instability\footnote{The $\sqrt{2}$
difference in the critical value of the temperature is due to a different normalization
of the gauge fields in \cite{Gladden:2024ssb} and in \eqref{model4}.}.
The ordered phase of the charged SYM plasma is however thermodynamically
unstable \cite{Gladden:2024ssb,Buchel:2025tjq}, and thus is
expected \cite{Gladden:2024ssb,Buchel:2005nt,Buchel:2010gd} to suffer
from the same charge clumping instabilities as that of the dual to
the RN black brane \eqref{rn}.

\underline{{\bf (Tr-III):}} Given \eqref{super}, the effective action \eqref{model4}
takes the form:
\begin{equation}
\begin{split}
&S_5=\frac{1}{16\pi G_5}\int_{\calm_5}d^5\xi\sqrt{-g} \biggl[
R-3 F_{\mu\nu} F^{\mu\nu}
-2\left(\del_\mu\phi\right)^2-\frac 12 \sinh^2(2\phi)\left(
\del_\mu\theta-3 A_\mu\right)^2
-4\calp
\biggr]\,,
\end{split}
\eqlabel{model5}
\end{equation}
where we introduced $\phi_4\equiv \phi$, $\theta_4\equiv \theta$, and $A_{\mu}^{(j)}\equiv A_\mu$.
Following \cite{Gubser:2009qm}, to have a properly normalized $R$-charge 2
superpotential of the boundary SYM we need to rescale the bulk gauge field as 
\begin{equation}
A_\mu = \frac 23 \hat{A}_\mu\,.
\eqlabel{rescale}
\end{equation}
Further introducing $\phi\equiv \frac \eta2$, the effective action \eqref{model5}
becomes identical to that of \cite{Gubser:2009qm}. The superconducting
phase transition in \cite{Gubser:2009qm} was reported at
\begin{equation}
\frac{T_c}{\hat \mu} = 0.060676(6)\,,
\eqlabel{gc}
\end{equation}
where $\hat \mu$ is the chemical potential of the $\hat A$ gauge field. Thus,
given \eqref{rescale}, the phase transition in the effective action \eqref{model5}
is at\footnote{We recovered this result directly from the effective action \eqref{model5}.}
\begin{equation}
\begin{split}
\frac{T_c}{\mu}\bigg|_{\rm superconducting} =&\ \frac 32\ \cdot\  0.060676(6)\ =\ 0.091014(9)\\
=&\ 0.57186(4)\ \cdot\ \frac{1}{2\pi} =\ 0.57186(4)\ \cdot\ \frac{T_c}{\mu}\bigg|_{\rm GIKS}\,,
\end{split}
\eqlabel{gc2}
\end{equation}
where we used \eqref{tc1}.

\section*{Acknowledgments}
Research at Perimeter Institute is supported by the Government of Canada through Industry
Canada and by the Province of Ontario through the Ministry of
Research \& Innovation. This work was further supported by
NSERC through the Discovery Grants program.

\bibliographystyle{JHEP}
\bibliography{n4rn3}

\end{document}